\documentclass{article}
\usepackage{amsfonts,amssymb,amsmath,amsthm,graphicx}
\usepackage{bm}      
\usepackage[english]{babel}

  \newtheorem{thm}{Theorem}
  \newtheorem{cor}{Corollary}

\let\phi=\varphi

\let\theta=\vartheta

\newcommand{\bC}{{\mathbb C}}

\newcommand{\bP}{{\mathbb P}}

\newcommand{\bZ}{{\mathbb Z}}

\DeclareMathOperator{\Un}{U}

\newcommand{\ket}{\rangle}

\begin{document}
\title{Projective Ring Line of a Specific Qudit}
\author{Hans Havlicek$^{1}$ and Metod Saniga$^{2}$\\
\\
\normalsize $^{1}$Institut f\" ur Diskrete Mathematik und Geometrie\\
\normalsize  Technische Universit\" at Wien, Wiedner Hauptstrasse 8-10\\
\normalsize  A-1040 Vienna, Austria\\
\normalsize  (havlicek@geometrie.tuwien.ac.at)
\\ \\
\normalsize  $^{2}$Astronomical Institute, Slovak Academy of Sciences\\
\normalsize  SK-05960 Tatransk\' a Lomnica, Slovak Republic\\
\normalsize  (msaniga@astro.sk)}

\date{}

\maketitle

\begin{abstract}\noindent
A very particular connection between the commutation relations of the elements of the generalized Pauli group of a
$d$-dimensional qudit, $d$ being a product of distinct primes, and the structure of the projective line over the (modular) ring $\bZ_{d}$ is
established, where the integer exponents of the generating shift ($X$) and clock ($Z$) operators are associated with submodules of  $\bZ^{2}_{d}$.
Under this correspondence, the set of operators commuting with a given one --- a perp-set --- represents a $\bZ_{d}$-submodule of $\bZ^{2}_{d}$.
A crucial novel feature here is that the operators are also represented by {\it non}-admissible pairs of $\bZ^{2}_{d}$. This additional degree of
freedom makes it possible to view any perp-set as a {\it set-theoretic} union of the corresponding points of the associated projective line.\\
\par\noindent
{\bf PACS Numbers:} 03.65.-a -- 03.65.Fd -- 02.10.Hh -- 02.40.Dr
\par\noindent
{\bf Keywords:} Qudit -- Generalized Pauli Group -- Projective Ring Line -- Commutation
                  Algebra of Generalized Pauli Operators

\end{abstract}

\section{Introduction}
The study of the finite-dimensional Hilbert spaces and their associated generalized Pauli operators has been a forefront issue of the quantum
information theory within the past few years. A substantial mathematical insight has been possible thanks to a number of novel graph-combinatorial
and algebraic geometrical concepts employed, see, e.\,g., \cite{ara}--\cite{psk} and references therein. Among the latter, it is the concept of a
projective line defined over a(n associative) ring with unity that acquired a distinguished footing \cite{psk}--\cite{qic}. In this approach, one
simply identifies the points of a projective ring line with the generalized Pauli operators (or the maximum commuting sets of them) pertaining to a
given Hilbert space and rephrases their commutation relations in terms of neighbour/distant relations between the points on the line in question.
Given this identification, it was possible to ``projective-ring-geometrize" any $N$-qubit Hilbert space \cite{psk}--\cite{adv},
two-qutrits \cite{spie,qic}, as well as to get important hints about the smallest composite case, viz. a six-dimensional Hilbert space \cite{pbs}.
A detailed examination of these particular cases led soon to a discovery of a more complex and unifying approach based on group-theoretical
considerations \cite{koen1,koen2}.
Adopting and properly generalizing the strategy pursued in the last two mentioned papers, we shall demonstrate, on the example of a specific
single qudit, that the concept of a projective ring line naturally emerges also in a context slightly different from that introduced and
elaborated in \cite{psk}--\cite{pbs}, with the finest traits of the structure of the projective line coming into play.

\section{The Pauli group $G$ of a single qudit}

Let $d>1$ be an integer and $\bZ_d:=\{0,1,\ldots,d-1\}$. Addition and
multiplication of elements from $\bZ_d$ will always be understood modulo $d$.

We consider the $d$-dimensional complex Hilbert space $\bC^d$ and denote by
\begin{equation*}
   \{\, |s\ket : s\in\bZ_d \}
\end{equation*}
a computational basis of $\bC^d$. Furthermore, let $\omega$ be fixed a
primitive $d$-th root of unity (e.\,g.,\ $\omega=\exp(2\pi i/d)$).

Now $X$ and $Z$ are unitary ``shift" and ``clock" operators on $\bC^d$ defined via $X|s\ket =
|s+1\ket$ and $Z|s\ket = \omega^s |s\ket$, respectively, for all $s\in\bZ_d$.
With respect to our computational basis the matrices of $X$ and $Z$ are
\begin{equation*}
    \begin{pmatrix}
    0      & 0     &\ldots &0      & 1\\
    1      & 0     &\ldots &0      & 0\\
    0      & 1     &\ldots &0      & 0\\
    \vdots &\vdots &\ddots &\vdots &\vdots\\
    0      & 0     &\ldots &1      & 0
    \end{pmatrix}
    \mbox{~~and~~}
    \begin{pmatrix}
    1      & 0     &0        &\ldots&  0\\
    0      &\omega &0        &\ldots&  0\\
    0      & 0     &\omega^2 &\ldots& 0\\
    \vdots &\vdots &\vdots   &\ddots &\vdots\\
    0  & 0 & 0     &\ldots   &\omega^{d-1}
    \end{pmatrix},
\end{equation*}
respectively.
\par
The subgroup of the unitary group $\Un_d$ generated by $X$ and $Z$, known by physicists as the (generalized) \emph{Pauli group}, will be
written as $G$. The operators $X^0=: I, X^1,\ldots,X^{d-1}$ form a cyclic
subgroup of $G$ with order $d$; the same properties hold for $Z^0,Z^1,\ldots,
Z^{d-1}$. Hence
\begin{equation}\label{eq:d-1}
    X^{d-1}=X^{-1} \mbox{~~and~~} Z^{d-1}=Z^{-1}.
\end{equation}
For all $s\in\bZ_d$ we have $XZ|s\ket = \omega^s|s+1\ket$ and $ZX|s\ket =
\omega^{s+1}|s+1\ket$. This gives the basic relation
\begin{equation}\label{eq:xzzx}
    \omega XZ = ZX.
\end{equation}
By virtue of (\ref{eq:d-1}) and (\ref{eq:xzzx}), each element of $G$, usually referred to as a (generalized) \emph{Pauli operator},  can be
written in the \emph{normal form}
\begin{equation}\label{eq:normalform}
    \omega^a X^b Z^c \mbox{~~for some integers~~}a,b,c\in \bZ_d.
\end{equation}
It is easy to see that this representation in normal form is \emph{unique}:
From $\omega^a X^b Z^c = \omega^{a'} X^{b'} Z^{c'}$ follows $\omega^{a-a'}
X^{b-b'} Z^{c-c'}=I$. As $|0\ket$ remains fixed under $Z^{c-c'}$ we obtain
$\omega^{a-a'} X^{b-b'}|0\ket=|0\ket$. This shows $b-b'=0$ and $a-a'=0$. Thus
$Z^{c-c'}=I$ which implies $c-c'=0$, as required. The uniqueness of the normal
form (\ref{eq:normalform}) will be crucial for our further exhibition.
\par
We immediately may read off from (\ref{eq:xzzx}) the following rule for
multiplication in $G$, when the factors are given in normal form:
\begin{equation*}
    (\omega^a X^b Z^c) (\omega^{a'} X^{b'} Z^{c'}) = \omega^{b'c + a+a'}
X^{b+b'} Z^{c+c'}.
\end{equation*}
Observe that the product is also in normal form. The term $b'c$ in the exponent
of $\omega$ on the right hand side shows that $G$ is a non-commutative group.
The uniqueness of the normal form implies also that $G$ is a group of order
$|G|=d^3$.
\par
The \emph{commutator\/}\footnote{We shall always be concerned with commutator
of operators in the sense of group theory. It must not be confused with the
commutator from ring theory which uses addition and multiplication of
operators.} of two operators $W$ and $W'$ is
\begin{equation}\label{eq:defcommutator}
    [W,W'] := W W'W^{-1}{W'} ^{-1}.
\end{equation}
If $W=\omega^a X^b Z^c$ and $W'=\omega^{a'} X^{b'}Z^{c'}$ are given in
normal form then it is immediate from (\ref{eq:xzzx}) that
\begin{equation}\label{eq:commutator}
    [\omega^a X^b Z^c, \omega^{a'} X^{b'}Z^{c'}] = \omega^{cb'-c'b}I.
\end{equation}
Recall that two operators commute if, and only if, their commutator (taken in
any order) is equal to $I$.
\par
We shall be concerned with two important normal subgroups of $G$:
\par
The \emph{centre\/} $Z(G)$ of $G$ is the set of all operators in $G$ which
commute with every operator in $G$. An operator $\omega^a X^b Z^c$ given in
normal form lies in $Z(G)$ precisely when (\ref{eq:commutator}) holds for any
choice of $a'$, $b'$, and $c'$. Setting $b':=0$, $c':=1$ we get $b=0$, whereas
$b':=1$ and $c':=0$ gives then $c=0$. These necessary conditions are also
sufficient, whence
\begin{equation*}\label{eq:centre}
    Z(G) = \{\omega^a I: a\in\bZ_d \}.
\end{equation*}
Note that $Z(G)$ is yet another cyclic subgroup of $G$ with order $d$.
\par
The \emph{commutator subgroup} $[G,G]$ is the smallest subgroup of $G$ which
contains all commutators $[W,W']$ with $W,W'\in G$. We follow the usual
convention to denote the commutator subgroup of $G$ by $G'$. From $[Z,X]=\omega
I$ follows that all powers of $\omega I$ are elements of $G'$. On the other
hand (\ref{eq:commutator}) shows that there are no other commutators but the
powers of $\omega I$. Altogether we obtain
\begin{equation}\label{eq:G'}
    G' = Z(G) = \{\omega^a I: a\in\bZ_d \}.
\end{equation}
It is easy to see from (\ref{eq:commutator}) that each element of $G'$ is
indeed a commutator, a property which need not be true for the commutator
subgroup of an arbitrary group.

\section{The ring associated with $G$}

By expressing the elements of our group $G$ in normal form we saw already that
several basic algebraic relations can be expressed solely in terms of the
exponents of $\omega$, $X$ and $Z$. These exponents are always elements of the
\emph{ring} $(\bZ_d,+,\cdot)$ of integers modulo $d$. To be more precise, this
ring is unital ($1b=b$ for all $b\in\bZ_{d}$) and commutative ($bc=cb$ for all
$b,c\in\bZ_d$). An element $b\in\bZ_d$ is a unit (an invertible element) if,
and only if $b$ and $d$ are coprime. If $d$ is a prime then every non-zero
element of $\bZ_d$ is invertible and $\bZ_d$ is a field, otherwise there are
non-invertible elements --- see, e.\,g., \cite{mcd,rag} for more details.
\par
We show now that the ring $\bZ_d$ ``lives'', up to isomorphism, also within our
group $G$. Let us consider the bijective mapping
\begin{equation*}
    \psi : \bZ_d \to G' : a \mapsto \omega^a I.
\end{equation*}
This is an isomorphism of the additive group $(\bZ_d,+)$ onto the
multiplicative group $(G',\cdot)$, since clearly
\begin{equation*}
    \psi(a+a')=\omega^{a+a'}I = \psi(a)\cdot\psi(a')
    \mbox{~~for all~~} a,a'\in\bZ_d.
\end{equation*}
However, in $\bZ_d$ we also have the binary operation of multiplication. We
obtain its counterpart in $G'$ via
\begin{equation*}
    \psi(aa')=\omega^{aa'}I = (\omega^a I)^{a'}=\psi(a)^{a'}
    \mbox{~~for all~~} a,a'\in\bZ_d.
\end{equation*}
Thus we \emph{could\/} use the bijection $\psi$ to turn $G'$ into an isomorphic
copy of the ring $(\bZ_d,+,\cdot)$ by defining ``new'' binary operations on $G$
in accordance with the two formulas from the above. However, we refrain from
doing so in order to avoid misunderstandings. (The ``new'' addition would be
the ``old'' multiplication.) It is nevertheless important to emphasise that such a
construction is possible.

\section{A symplectic module associated with $G$ and the commutation algebra of Pauli operators}
As $(G,\cdot)$ is a non-commutative group, it cannot be isomorphic to the
additive group of any module. Recall that the factor group of $G$ by any normal
subgroup is commutative if, and only if, this normal subgroup contains the
commutator subgroup $G'$. This means that the ``largest'' commutative group we
can obtain from $G$ by factorisation is the factor group
\begin{equation*}\label{eq:factorgroup}
    G/G'.
\end{equation*}
Taking into account our normal form (\ref{eq:normalform}) and the description
of $G'$ in (\ref{eq:G'}), the group $G/G'$ comprises all cosets
\begin{equation}\label{normalform'}
    G'X^bZ^c \mbox{~~where~~}b,c\in\bZ_d.
\end{equation}
Each element of $G/G'$ can be written in a \emph{unique} way in this
\emph{normal form}. As a by-product of this uniqueness, we learn from
(\ref{normalform'}) that the factor group $G/G'$ has order $d^2$.
Multiplication in $G/G'$ is governed by the formula
\begin{equation}\label{eq:mult'}
    (G'X^bZ^c)(G'X^{b'}Z^{c'}) = G'X^{b+b'} Z^{c+c'}
    \mbox{~~for all~~} b,c,b',c'\in\bZ_d.
\end{equation}
\par
Let us consider the bijective mapping
\begin{equation*}
    \phi : \bZ_d^2 \to G/G' : (b,c) \mapsto G'X^bZ^c.
\end{equation*}
Note that the elements of $\bZ_d^2$ are written as rows. Sometimes they will be
called \emph{vectors}. We now consider $(\bZ_d^2,+)$ as a commutative group
with the addition ($+$) defined componentwise. Then (\ref{eq:mult'})
establishes immediately that $\phi$ is an isomorphism of the additive group
$(\bZ_d^2,+)$ onto the multiplicative group $(G/G',\cdot)$.
\par
But $\bZ_d^2$ is also a module over $\bZ_d$ in the usual way. Thus we
\emph{could\/} use the bijections $\psi:\bZ_d\to G'$ and $\phi:\bZ_d^2\to G/G'$
to turn $G/G'$ into an isomorphic module over $G'$. Like before, it is worth
noting that this is possible, but the actual construction will not be needed.
Let us just present an example: Given $a,b,c\in \bZ_d$ we have on the one hand
$a(b,c)=(ab,ac)$. On the other hand the ``product'' of the ``scalar'' $\omega^a
I\in G'$ with the ``vector'' $G'X^b Z^c$ would equal the ``vector''
$G'X^{ab}Z^{ac}$.
\par
Recall that our main goal is to describe whether or not two operators of $G$
commute. Since $G/G'$ is a commutative group, any information of this kind is
eliminated by our passage from $G$ to the factor group $G/G'$. This is why in
the following construction we use not only the group $G/G'$, but also the group
$G$ and the commutator subgroup $G'$:
\par
Let $G'X^bZ^c$ and $G'X^{b'}Z^{c'}$ be elements of $G/G'$ in normal form. We
associate with them the commutator
\begin{equation*}\label{eq:lift}
    [X^bZ^c, X^{b'}Z^{c'}] = \omega^{cb'-c'b}I\in G'.
\end{equation*}
This assignment uses the group $G$. It is independent of the choice of
representatives from the cosets $G'X^bZ^c$ and $G'X^{b'}Z^{c'}$, since $a$ and
$a'$ do not appear on the right hand side of (\ref{eq:commutator}).
\par
By virtue of the bijections $\phi^{-1}:G/G'\to\bZ_d^2$ and $\psi^{-1}:G'\to\bZ_d$
we are now in a position to transfer this construction to our $\bZ_d$-module
$\bZ_d^2$. This gives a mapping\footnote{Of course the symbol $[\cdot,\cdot]$
has two different meanings in (\ref{eq:defcommutator}) and
(\ref{eq:alternating}).}
\begin{equation}\label{eq:alternating}
    [\cdot,\cdot] : \bZ_d^2 \to\bZ_d : \big((b,c),(b',c')\big) \mapsto cb'-c'b
\end{equation}
which just describes the commutator of two elements of $G$ (given in normal
form) in terms of our $\bZ_d$-module. There are several ways to rewrite the
mapping (\ref{eq:alternating}), for example
\begin{equation}\label{eq:determinant}
    \big[(b,c),(b',c')\big]
    = (b,c)\begin{pmatrix} 0 &-1\\1&\hphantom{-}0\end{pmatrix}\begin{pmatrix}b'\\c'\end{pmatrix}
    = \det \begin{pmatrix} b' &c'\\b\hphantom{'}&c\hphantom{'}\end{pmatrix}.
\end{equation}
By this formula, the mapping $[\cdot,\cdot]$ is a \emph{bilinear form\/} on
$\bZ_d^2$. Clearly, this form is alternating, i.~e., $\big[(b,c),(b,c)\big]=0$
for all $(b,c)\in\bZ_d^2$. As usual, we write $(b,c)\perp(b',c')$ if
$\big[(b,c),(b',c')\big]=0$ and speak of \emph{orthogonal} (or: \emph{perpendicular}) vectors (with
respect to $[\cdot,\cdot]$). As an alternating bilinear form is always skew
symmetric, our orthogonality of vectors is a symmetric relation.
As our form $[\cdot,\cdot]$ is \emph{non-degenerate}, i.\,e., only the zero-vector is orthogonal to all other vectors 
of $\bZ_d^2$, we have indeed a \emph{symplectic module}.
\par
Summing up, we see that the set of operators in $G$ which commute with a fixed
operator $\omega^aX^bZ^c$ corresponds to the \emph{perpendicular set} (shortly
the \emph{perp-set}) of $(b,c)$, viz.\
\begin{equation*}
    (b,c)^\perp := \big\{(u,v)\in\bZ_d^2 : (b,c)\perp (u,v) \big\}.
\end{equation*}
The perp-set of $(b,c)$ is closed under addition and multiplication by ring
elements. Also, it is non empty, since
\begin{equation}\label{eq:perpset}
    \bZ_d(b,c)\subset(b,c)^\perp.
\end{equation}
So, $(b,c)^\perp$ is a $\bZ_d$-submodule of $\bZ_d^2$. We shall exhibit
perp-sets in detail in the following sections.

\section{The projective line over $\bZ_d$ and the commutation algebra of Pauli operators}
In order to say more about perp-sets in $\bZ_d^2$ we shall use some basic facts
about the projective line over the ring $\bZ_d$. We do not need the theory of
projective ring lines in its most general form here, since our ring $\bZ_d$ is
commutative and finite. This will allow to work with determinants and state
some definitions in a simpler way. While we sketch here some basic notions
and results, the reader is referred to \cite{bh}--\cite{blhr} for further details and proofs.
\par
First, let us consider any vector $(b,c)\in\bZ_d^2$. It generates the cyclic
submodule
\begin{equation*}
    \bZ_d(b,c) = \{(ub,uc):u\in\bZ_d\}
\end{equation*}
Such a cyclic submodule is called \emph{free\/}, if the mapping
$u\mapsto(ub,uc)$ is injective. In this case the vector (or: pair) $(b,c)$ is
called \emph{admissible}. Any free cyclic submodule of $\bZ_d^2$ has precisely
$d$ vectors, including the zero-vector. However, not all vectors $\neq(0,0)$ of
a free cyclic submodule need to be admissible. If $(b,c)$ is an admissible vector
then $(ub,uc)$ is also admissible if, and only if, $u\in\bZ_d$ is an invertible
element. Thus, if $d$ is not a prime each free cyclic submodule of $\bZ_d^2$
contains at least one non-admissible vector other than $(0,0)$.
\par
For our ring $\bZ_d$ there are several other ways of describing admissible
vectors, as the following assertions are equivalent for any vector
$(b,c)\in\bZ_d^2$:
\begin{enumerate}
\item
The vector $(b,c)$ is \emph{unimodular}, i.~e., there exist elements
$u,v\in\bZ_d$ with
\begin{equation*}
    ub + vc = 1.
\end{equation*}
\item The vector $(b,c)$ is the first row of an invertible $2\times 2$
matrix with entries in $\bZ_d$.
\item
The vector $(b,c)$ is the first vector\footnote{All bases of $\bZ_d^2$ consist
of two admissible vectors. In general, a module over a ring may have bases of
different size.} of a basis of $\bZ_d^2$. (This means that there is such a
vector $(b',c')\in\bZ_d^2$ that the mapping
\begin{equation*}
    \bZ_d^2\to\bZ_d^2:(u,u')\mapsto u(b,c)+u'(b',c')
\end{equation*}
is a bijection.)
\end{enumerate}
\par
In a more geometric language, motivated by classical analytic projective
geometry over the real or complex numbers, a free cyclic submodule of $\bZ_d^2$
is called a \emph{point}. The point set
\begin{equation*}\label{eq:projectiveline}
    \bP_1(\bZ_d):=\{\bZ_d(c,d) : (c,d) \mbox{~is admissible} \}
\end{equation*}
is the \emph{projective line\/} over the ring $\bZ_d$. According to this
definition a point is a set of vectors. In ``genuine'' projective geometry over
a ring the individual vectors contained in a point are of no particular
interest. They are merely a useful tool for doing geometry in terms of
coordinates. For us, however, the vectors within a point will be significant.
This is of course in sharp contrast to Euclid's point of view: \emph{A point is
that which has no part.}
\par
Two points $\bZ_{d}(b,c)$ and $\bZ_{d}(b',c')$ of $\bP_1(\bZ_{d})$ are called
\emph{distant} if $(b,c),(b',c')$ is a basis of $\bZ_d^2$. Two distant points
share only the zero vector $(0,0)$. Otherwise, the points are called
\emph{neighbouring}. Thus, two neighbouring points have always a non-zero
vector in common.
\par
We are now in a position to state a first, preliminary result about perp-sets.
\begin{thm}\label{thm:weakresult}
Let $(b,c)\in\bZ_d^2$ be any vector and let $\bZ_d(b',c')$ be any point of the
projective line $\bP_1(\bZ_d)$ which contains the vector $(b,c)$. Then the
following assertions hold:
\begin{enumerate}
\item The point $\bZ_d(b',c')$ is a subset of the perp-set $(b,c)^\perp$.
\item Under the additional assumption that $\bZ_d(b,c)$ is also a point, we have
\begin{equation*}\label{eq:admissible}
    (b,c)^\perp=\bZ_d(b,c)=\bZ_d(b',c').
\end{equation*}
\end{enumerate}
\end{thm}
\begin{proof}
Ad (a): By (\ref{eq:perpset}) and the assumption of the theorem,
$(b,c)\in\bZ_d(b',c')\subset(b',c')^\perp$. We infer from the symmetry of the
relation $\perp$ that $(b',c')\in (b,c)^\perp$. Also, since $(b,c)^\perp$ is a
submodule, we obtain that the entire point $\bZ_d(b',c')$ is a subset of
$(b,c)^\perp$.
\par
Ad (b): As $(b,c)$ is a unimodular vector, there exists a pair $(\widetilde
c,-\widetilde b)\in\bZ_{d}^{2}$ such that $b\widetilde c - c\widetilde b=1$. This means
\begin{equation*}
    \det\begin{pmatrix}
            b & c \\\widetilde b & \widetilde c
        \end{pmatrix} = 1
\end{equation*}
which in turn tells us that $(b,c)$ and $(\tilde b, \tilde c)$ form a basis of $\bZ_d^2$.
Each vector $(u,v)\in\bZ_d^2$ can be expressed in a unique way as a linear
combination
\begin{equation*}
    (u,v) = w(b,c)+\widetilde w(\widetilde b,\widetilde c)
    \mbox{~~with~~}w,\widetilde w\in\bZ_d.
\end{equation*}
By (\ref{eq:determinant}), a necessary and sufficient condition for $(u,v)$ to
lie in $(b,c)^\perp$ reads
\begin{equation*}
    \det\begin{pmatrix}
    wb+\widetilde w\widetilde b &wc+\widetilde w\widetilde c
    \\
    b & c
    \end{pmatrix}
    = \widetilde w (\widetilde b c - b\widetilde c) = - \widetilde w = 0.
\end{equation*}
Therefore $(b,c)^\perp = \bZ_d(b,c)$. Finally, we infer from
$(b,c)\in\bZ_d(b',c')$ that the point $\bZ_d(b,c)$ is a subset of the point
$\bZ_d(b',c')$. These points coincide, as both have precisely $d$ vectors.
\end{proof}
Let us give an example, where $d=6$. We consider the vector $(2,0)$ which
cannot be unimodular, because $2b'+0c'=1$ has no solution in $\bZ_6$. There are
only three distinct multiples of $(2,0)$, namely $(0,0)$, $(2,0)$, and $(4,0)$.
This indicates once more that $(2,0)$ is not unimodular (or: admissible). We
infer from
\begin{equation*}
    (2,0) = 4 (5,0)  = 4(2,3) = 4(5,3)
\end{equation*}
that there are (at least) three points containing $(2,0)$. The subsequent
remarks are immediate from Theorem \ref{thm:betterresult} which will be
established below. However, their verification is also an easy exercise which
can be carried out without any background knowledge: The projective line
$\bP_1(\bZ_6)$ has precisely twelve points. There are no other points
containing $(2,0)$ than those mentioned before. The perp-set of $(2,0)$
coincides with the set-theoretic union of those three points, hence
\begin{eqnarray*}
    (2,0)^\perp &=& \bZ_6(5,0)\cup \bZ_6(2,3)\cup \bZ_6(5,3)\\
    &=&             \{(5,0),(4,0),(3,0),(2,0),(1,0),(0,0),\\
    && \hphantom{\{}  (2,3),(4,0),(0,3),(2,0),(4,3),(0,0),\\
    && \hphantom{\{}  (5,3),(4,0),(3,3),(2,0),(1,3),(0,0) \}.
\end{eqnarray*}
This is a set of $18-6=12$ vectors, because $(2,0)$, $(4,0)$ and $(0,0)$ are
vectors which belong to all three points.

\section{A particular case: $d$ is square-free}
While Theorem \ref{thm:weakresult} describes the perp-set of any admissible
vector, the result for non-admissible vectors is unsatisfactory. The aim of
this section is to improve the results of Theorem \ref{thm:weakresult} under
the additional hypothesis that the number $d$ is square-free. Throughout this
section we adopt the assumption that
\begin{equation}\label{eq:factors}
    d = p_1p_2\cdots p_r,
\end{equation}
where $p_1,p_2,\ldots,p_r$ are $r\geq 1$ distinct prime numbers. The ring
$\bZ_d$ is isomorphic to the outer direct product
\begin{equation}\label{eq:outer}
    \bZ_{p_1}\times\bZ_{p_2}\times\cdots\times\bZ_{p_r}
\end{equation}
of $r$ finite \emph{fields}. Let us recall how this isomorphism arises: We
consider the ring elements
\begin{equation*}\label{}
    q_k:=p_1\cdots p_{k-1}p_{k+1}\cdots p_{r}, \mbox{~~where~~} k\in\{1,2,\ldots,r\}.
\end{equation*}
(For $r=1$ this product is empty, whence $q_1=1$.) The ring $\bZ_d$ is the
inner direct product of the principal ideals
\begin{equation*}\label{}
    J^{(k)}:=\bZ_d q_k \mbox{~~where~~} k\in\{1,2,\ldots,r\}.
\end{equation*}
Given any element $y\in\bZ_d$ there exists a unique decomposition
\begin{equation*}\label{}
    y = y^{(1)}+y^{(2)}+\cdots+y^{(r)} \mbox{~~with~~} y^{(k)} \in J^{(k)}.
\end{equation*}
We refer to the elements $y^{(k)}$ as the \emph{components} of $y$. In terms of
this decomposition we can add and multiply elements of $x,y\in\bZ_d$
componentwise, i.~e.
\begin{equation*}
    (x+y)^{(k)} =x^{(k)}+y^{(k)} \mbox{~~and~~} (x\cdot y)^{(k)} =x^{(k)}\cdot
    y^{(k)}.
\end{equation*}
In particular, the unit element $1\in\bZ_d$ has the decomposition
\begin{equation*}\label{eq:representation1}
    1 = 1^{(1)}+1^{(2)}+\cdots+1^{(r)}.
\end{equation*}
For each $k\in\{1,2,\ldots,r\}$ the ideal $J^{(k)}$ is a \emph{field\/}
isomorphic to $\bZ_{p_k}$. There is only one isomorphism $J^{(k)}\to
\bZ_{p_k}$; it takes the element $1^{(k)}$ to the unit element $1\in\bZ_{p_k}$.
Note that each element of $J^{(k)}$ can be written as
$1^{(k)}+1^{(k)}+\cdots+1^{(k)}$ with a finite number of summands. Then its
isomorphic image in $\bZ_{p_k}$ is the sum $1+1+\cdots+1$ with the same number
of summands. Below we shall always use the representation of $\bZ_d$ as the
inner direct product of the ideals $J^{(k)}$ rather than the isomorphic model
given in (\ref{eq:outer}).
\par
As a first application we obtain the following characterisation: An element
$y\in\bZ_d$ is invertible if, and only if, all its components are non-zero. In
this case the $k$-th component of the element $y^{-1}$ is the unique solution
in $J^{(k)}$ of the equation $y^{(k)}x=1^{(k)}$ in the unknown $x$. Therefore
the number of invertible elements in $\bZ_d$ is
\begin{equation}\label{eq:invertible}
    \prod_{k=1}^{r}(p_k-1).
\end{equation}
A similar description holds for the points of $\bP_1(\bZ_d)$: A pair $(b,c)$ is
unimodular (or: admissible) if, and only if, there exist elements $u,v\in\bZ_d$
with
\begin{equation*}\label{eq:unimodular_k}
    u^{(k)}b^{(k)}+v^{(k)}c^{(k)}=1^{(k)}\mbox{~~for all~~}
    k\in\{1,2,\ldots,r\}.
\end{equation*}
Since each ideal $J^{(k)}$ is isomorphic to a field, the last equation is
equivalent to
\begin{equation}\label{eq:nonzero_k}
    (b^{(k)},c^{(k)})\neq (0,0)\mbox{~~for all~~}k\in\{1,2,\ldots,r\}.
\end{equation}
\par
We are now in a position to state our main result. Note that the
set-theoretic union of points gives a set of vectors.
\par
\begin{thm}\label{thm:betterresult}
Let the square-free integer $d > 1$ be given as in \emph{(\ref{eq:factors})}.
Also, let $(b,c)\in\bZ_d^2$. We denote by $K$ the set of those indices
$k\in\{1,2,\ldots,r\}$ such that $(b^{(k)},c^{(k)})=(0,0)$. Then the following
hold:
\begin{enumerate}
\item
The vector $(b,c)$ is contained in precisely
\begin{equation}\label{eq:points}
    \prod_{k\in K} (p_k+1)
\end{equation}
points of $\bP_1(\bZ_d)$.
\item
The set-theoretic union of these points equals the perpendicular set of the
vector $(b,c)$.
\item
The perpendicular set of the vector $(b,c)$ satisfies
\begin{equation}\label{eq:perpvectors}
   |(b,c)^\perp| = d\prod_{k\in K}p_k.
\end{equation}
\end{enumerate}
\end{thm}
\begin{proof}
Ad (a): First, let us determine all admissible vectors $(b',c')\in\bZ_{d}^{2}$ such
that $(b,c)=u(b',c')$ for some $u\in\bZ_d$. So
\begin{equation}\label{eq:offK}
    (b^{(j)},c^{(j)})=u^{(j)}({b'}{}^{(j)},{c'}{}^{(j)})\neq(0,0)
    \mbox{~~for all~~}j\in\{1,2,\ldots,r\}\setminus K
\end{equation}
and
\begin{equation}\label{eq:inK}
    (b^{(k)},c^{(k)})=u^{(k)}({b'}{}^{(k)},{c'}{}^{(k)})=(0,0)\neq ({b'}{}^{(k)},{c'}{}^{(k)})
    \mbox{~~for all~~}k\in K.
\end{equation}
We obtain $u^{(j)}\neq 0$ from (\ref{eq:offK}), whence
$({b'}{}^{(j)},{c'}{}^{(j)})$ is one of the $p_j-1$ distinct multiples of
$(b^{(j)},c^{(j)})$ by a non-zero factor in $J^{(j)}$. Next, (\ref{eq:inK})
implies $u^{(k)}=0$, whence $({b'}{}^{(k)},{c'}{}^{(k)})$ is one of the
$p_k^2-1$ non-zero pairs with entries from $J^{(k)}$. These necessary
conditions are also sufficient so that we obtain
\begin{equation*}\label{eq:total}
    \prod_{j\notin K} (p_j-1) \prod_{k\in K} (p_k+1)(p_k-1)
\end{equation*}
admissible vectors with the required property. Dividing by the number of
invertible elements of $\bZ_d$, as stated in (\ref{eq:invertible}), gives the
number of points containing the vector $(b,c)$.
\par
Ad (b): By Theorem \ref{thm:weakresult} (a), each point containing $(b,c)$ is a
subset of $(b,c)^\perp$. So the same property holds for the union of all these
points. The proof will be accomplished by showing that for any vector $(x,y)\in
(b,c)^\perp$ there is a point $\bZ_d(b',c')\in\bP_1(\bZ_d)$ which has $(x,y)$
and $(b,c)$ among its vectors. We define $b'$ and $c'$ in terms of their
components as follows: For all $j\in\{1,2,\ldots,r\}\setminus K$ we let
$({b'}{}^{(j)},{c'}{}^{(j)}) := ({b}^{(j)},{c}^{(j)})$. For the remaining
indices $k\in K$ we define
\begin{equation*}\label{}
    ({b'}{}^{(k)},{c'}{}^{(k)}) :=
    \left\{\begin{array}{ll}
    (x^{(k)},y^{(k)}) & \mbox{if~}(x^{(k)},y^{(k)}) \neq (0,0),\\
    (1^{(k)},1^{(k)}) & \mbox{otherwise.}
    \end{array}\right.
\end{equation*}
According to our definition and (\ref{eq:nonzero_k}) the submodule
$\bZ_d(b',c')$ is a point. Letting $u^{(j)}:=1^{(j)}$ for all $j\notin K$ and
$u^{(k)}:=0$ for all $k\in K$ yields $(b,c) = u(b',c')$.

\par
Finally, we establish the existence of an element $v\in\bZ_d$ with
$(x,y)=v(b',c')$. For this purpose the components of $v$ can be chosen as
follows: If $j\notin K$ then $(x,y)\perp(b,c)$ together with
(\ref{eq:determinant}) gives
\begin{equation*}\label{eq:det_j}
    \det \begin{pmatrix} x^{(j)} &y^{(j)}\\b^{(j)}&c^{(j)}\end{pmatrix}=0.
\end{equation*}
As this is a determinant over the field $J^{(j)}$, and because the second row
is non-zero, we can define $v^{(j)}\in J^{(j)}$ via
$(x^{(j)},y^{(j)})=v^{(j)}(b^{(j)},c^{(j)})$. If $k\in K$ and
$(x^{(k)},y^{(k)}) \neq (0,0)$ we set $v^{(k)}:=1^{(k)}$, otherwise we let
$v^{(k)}:=0$.
\par
Ad (c): By (b), it suffices to count the number of vectors $(b'',c'')$ which
are a multiple of an admissible vector as described in part (a) of the present
proof. For each $j\notin K$ the pair $(b''{}^{(j)},c''{}^{(j)})$ can be chosen
as any of the $p_j$ multiples of $(b^{(j)},c^{(j)})$ by a factor in $J^{(j)}$,
whereas for each $k\in K$ the pair $(b''{}^{(k)},c''{}^{(k)})$ can be chosen
arbitrarily in $p_k^2$ ways. Hence there are
\begin{equation*}
    \prod_{j\notin K}p_j\cdot\prod_{k\in K}p_k^2 = d\prod_{k\in K}p_k
\end{equation*}
such vectors.
\end{proof}
As a by-product of Theorem \ref{thm:betterresult} we may infer from
(\ref{eq:points}) that the projective line $\bP_1(\bZ_d)$ has precisely
$\prod_{k=1}^{r} (p_k+1)$ points. Also, returning to our initial problem, we
obtain the following result.

\begin{cor}
With the settings and notations of Theorem \emph{2} the number of
operators in the generalized Pauli group $G$ which commute with the operator $\omega^aX^bZ^c\in G$ equals the value given
by \emph{(\ref{eq:perpvectors})} multiplied by d.
\end{cor}

\section{Conclusion}
Given the generalized Pauli group associated with a $d$-dimensional qudit, $d$ a product
of distinct primes, a general formula was derived for the number of generalized Pauli operators
commuting with a given one. This formula is based on the properties of (sub)modules of the associated
modular ring $\bZ_d$ and finds its natural interpretation in the properties of the 
projective line defined over $\bZ_d$. When compared with other works on the subject \cite{psk}--\cite{pbs}, our approach 
makes also use of {\it non}-admissible pairs of elements of the ring in question, thereby giving the physical meaning
to the full structure of the line; moreover, it seems to be readily generalizable to tackle the case where $d$
also contains powers of primes.

\section*{Acknowledgements}
This work was supported by the Science and
Technology Assistance Agency under the contract $\#$
APVT--51--012704, the VEGA grant agency projects $\#$ 2/6070/26
and $\#$ 7012 and by
the $\langle$Action Austria--Slovakia$\rangle$ project $\#$ 58s2 ``Finite Geometries Behind Hilbert Spaces."

\end{document}